\documentstyle[12pt,epsf]{article}

\begin{document}

\begin{titlepage}

\begin{flushright}
\end{flushright}

\vskip 2.5cm

\begin{center}
{\Large \bf Spontaneous Parametric Down-Conversion
Experiment to Measure Both Photon Trajectories
and Double-Slit Interference}
\end{center}

\vspace{1ex}

\begin{center}
{\large M. S. Altschul$^{1}$ and B. Altschul$^{2}$}

\vspace{5mm}
{$^{1}$ \sl Salem, OR, 97306 USA} \\
{\tt Martin.Altschul@kp.org} \\

\vspace{5mm}
{$^{2}$ \sl Department of Mathematics} \\
{\sl Massachusetts Institute of Technology} \\
{\sl Cambridge, MA, 02139 USA} \\
{\tt baltschu@mit.edu} \\

\end{center}

\vspace{2.5ex}

\medskip

\centerline {\bf Abstract}

\bigskip

Recent work using spontaneous parametric down-conversion (SPDC) has made
possible investigations of the Einstein-Podolsky-Rosen paradox in its original
(pos\-i\-tion-mo\-men\-tum) form. We propose an experiment that uses SPDC
photon pairs to measure through which slit a photon passes while simultaneously
observing double-slit interference.

\bigskip 

\end{titlepage}

\newpage

Recently, there has been increasingly sophisticated use of spontaneous
parametric down-conversion (SPDC) to elucidate Einstein-Podolsky-Rosen-type
state entanglements. In general, these experiments take the following
form~\cite{ref-zeilinger}. When
down-conversion occurs, conservation of momentum requires that the signal-idler
pairs propagate in strongly correlated directions that are in some sense
``opposite''~\cite{ref-pittman,ref-jost}. The pairs are emitted from two
possible activations regions ($A$ and $B$) in a suitable crystal. The signal
photons are subjected to a double slit, in which the slits correspond to $A$
and $B$. This process produces double-slit interference if and only if it is
subjected to coincidence counting with an $A+B$ idler beam, a phenomenon often
called quantum erasure. In an elegant example~\cite{ref-kim}, there are two
$A+B$ idler beams with different phase factors, producing two double-slit
coincidence patterns whose peaks and troughs are perfectly out of phase, and
therefore sum to a perfectly non-double-slit pattern. In order that experiments
of this type should support the standard interpretation of quantum mechanics,
it is essential that the sum total pattern not display double-slit
interference.

Srikanth~\cite{ref-srikanth1,ref-srikanth2} has proposed a variant of this type
of experiment in which the double-slit observer may observe interference
without coincidence counting, a prospect which threatens the standard
interpretation. It is our intention to boil this type of experiment down to a
few essentials and see whether this threat really exists.

At the core of the experiment is the necessary condition for resolution of a
double-slit pattern,
\begin{equation}
\label{eq-distineq}
\frac{w}{d}<\frac{\lambda}{s},
\end{equation}
where
\begin{eqnarray*}
w & = & {\rm width\, of\, the\, source,}\\
d & = & {\rm distance\, from\, the\, source\, to\, the\, slits,\, and}\\
s & = & {\rm slit\, separation}.
\end{eqnarray*}
For strongly correlated photon pairs, we expect that pair detection varies as
\begin{equation}
\sqrt{2\pi}\frac{\phi}{\phi_{0}}\exp\left[-\frac{1}{2}\left(\frac{\phi}{\phi
_{0}}\right)^{2}\right],
\end{equation}
where $\phi$ is the angular deviation from perfect opposition, and $\phi_{0}$
is the one-dimensional standard deviation. It is clear that the experiment
harbors a potentially interesting result only if double-slit interference can
be shown to occur when
\begin{equation}
\label{eq-phiineq}
\frac{s}{d}>\phi_{0}.
\end{equation}
So, from~(\ref{eq-distineq}) and~(\ref{eq-phiineq}),
\begin{eqnarray}
\frac{\lambda}{w} & > & \frac{s}{d}>\phi_{0}\nonumber\\
\label{eq-wineq}
\frac{\lambda}{\phi_{0}} & > & w.
\end{eqnarray}
Using condition~(\ref{eq-wineq}) and various SPDC
results~\cite{ref-pittman,ref-jost,ref-kim}, we shall construct an experiment
in which
\begin{eqnarray*}
\lambda & = & {\rm 702\, nm,\, and}\\
\phi_{0} & = & {\rm 2\, mrad.}
\end{eqnarray*}

If we set $\frac{s}{d}=6\phi_{0}=12$ mrad to achieve separation between slits
$A$ and $B$, Equation~\ref{eq-wineq} is replaced by
\begin{equation}
w<\frac{\lambda}{6\phi_{0}}=(83.3)\lambda,
\end{equation}
We then set $w = 25\lambda$ to make the interference pattern easily resolvable.
Then if we set $d=600$ mm, we have
\begin{eqnarray*}
w & = & 1.75\times 10^{-2}\,{\rm mm,\, and}\\
s & = & 7.2\,{\rm mm,}
\end{eqnarray*}
and the experiment is summarized by Figure~1. Note that because of
the small value of $\frac{s}{d}$, we have for the sake of clarity used
different scales for the horizontal and vertical axes.

The distances $A'A''$ and $B'B''$ are 1.46 mm each, so we define
$x\equiv A'C=B'D=\frac{1}{2}A'A''=$ 0.73 mm.
0.73 mm is close to the usual value for the extent of the activation area (and
therefore the coherence length). We therefore choose to work with a crystal (or
perhaps just with an activation area) that is 0.73 mm thick. We see from
Figure~1 that $\frac{3}{5}$ of the points that have access to the slits
have access to both slits (the shaded area). Therefore, the naive conclusion is
that $\frac{3}{4}$ of the counts beyond the slits are double-slit counts. The
total pattern {\em without} coincidence counting should be predominantly a
double-slit pattern.

We complete the experiment by placing detectors $\bar{A}$ and $\bar{B}$
``opposite'' slits $A$
and $B$, at a distance $d'\gg d$, so that $\frac{w}{d'}$ is negligible. The
detectors have radius $2\phi_{0}$.

It is then difficult to escape the conclusion that coincidence counting with
either $\bar{A}$ or $\bar{B}$ alone will include a significant double-slit
component. Note that
a time delay for the idlers can be inserted, so that the double-slit counts are
registered before $\bar{A}$ and $\bar{B}$ are activated.

The one loophole that could avoid this conclusion and still maintain the
standard interpretation of quantum mechanics would be if the photon wave
functions are composed primarily of waves that appear to radiate from points in
zones $A$ and $B$ that are outside (the activation area of) the crystal.

Note that
\begin{equation}
K_{pe}\equiv\frac{x}{\lambda}=1.04\times 10^{3}
\end{equation}
is a measure of the system's {\em potential entanglement}, while
\begin{equation}
K_{ae}\equiv\frac{\pi/2}{\phi_{0}}=0.79\times 10^{3}
\end{equation}
is a measurement of the {\em actual entanglement}. We have designed the
experiment so that
\begin{equation}
0.5<\frac{K_{ae}}{K_{pe}}<1.
\end{equation}

We can determine more generally the conditions for observing the desired
effect. We define three parameters $f$, $g$, and $h$ be replacing three
inequalities with equations.
\begin{displaymath}
K_{pe}>K_{ae},
\end{displaymath}
for degree of entanglement, becomes
\begin{displaymath}
K_{pe}=fK_{ae};
\end{displaymath}
\begin{displaymath}
\frac{s}{d}>\phi_{0},
\end{displaymath}
for discriminating A from B, becomes
\begin{displaymath}
\frac{s}{d}=4g\phi_{0};
\end{displaymath}
and
\begin{displaymath}
\frac{\lambda}{s}>\frac{w}{d},
\end{displaymath}
for double-slit pattern resolution, becomes
\begin{displaymath}
\frac{\lambda}{s}=2h\frac{w}{d}.
\end{displaymath}
We then find that
\begin{equation}
\phi_{0}=\frac{1}{32\pi(fg^{2}h)}
\end{equation}
and
\begin{equation}
w=\frac{\lambda}{8gh\phi_{0}}.
\end{equation}
It appears that the threshold for observing the effect is about $f\approx1$,
$g\approx1$, $h\approx1$ with
\begin{eqnarray*}
\phi_{0} & = & 10\, {\rm mrad,\, and}\\
w & = & (12.5)\lambda,
\end{eqnarray*}
so even in the threshold case, diffraction by the aperture is not a critically
limiting factor. Current SPDC results allow us to propose a configuration in
which
\begin{eqnarray*}
fg^{2}h & \approx & 5\\
hg^{2} & = & 3.75,
\end{eqnarray*}
with $\phi_{0}=2$ mrad. If we fix $f$ then $g^{2}h=\frac{1}{132\phi_{0}}$.
$g^{2}h$ is a visibility parameter for the effect, which improves if the value
of $\phi_{0}$ decreases.

We conclude that the experiment is technically viable and consistent, and that
an interesting result is expected, provided that the stated loophole does not
apply.

\pagebreak

\begin{figure}
\label{fig}
\caption{A schematic of the experiment.}
\end{figure}

\end{document}